# SIGMA WEB INTERFACE FOR REACTOR DATA APPLICATIONS

**B. Pritychenko* and A.A. Sonzogni**
National Nuclear Data Center
Brookhaven National Laboratory
Upton, NY 11973-5000, USA
pritychenko@bnl.gov; sonzogni@bnl.gov

## ABSTRACT

We present Sigma Web interface which provides user-friendly access for online analysis and plotting of the evaluated and experimental nuclear reaction data stored in the ENDF-6 and EXFOR formats. The interface includes advanced browsing and search capabilities, interactive plots of cross sections, angular distributions and spectra, nubars, comparisons between evaluated and experimental data, computations for cross section data sets, pre-calculated integral quantities, neutron cross section uncertainties plots and visualization of covariance matrices. Sigma is publicly available at the National Nuclear Data Center website at *http://www.nndc.bnl.gov/sigma*.

## 1. INTRODUCTION

Increasing energy demand, concerns over climate change and high volatility of petroleum market make a very strong case for nuclear power renaissance in the US and worldwide. New power reactor units will be built utilizing the latest technologies and their design will incorporate the best available nuclear data. It is essential, in parallel with the data developments, to produce effective data analysis and processing tools for reactor physicists and engineers utilizing the modern computer technologies.

Recent work on the release of ENDF/B-VII.0 evaluated nuclear reaction data library by the CSEWG collaboration [1] motivated the development of an advanced Sigma Web interface for data dissemination, online analysis and processing of ENDF-6 formatted libraries [2]. This interface is based on Java Web and database technologies and currently provides easy access, advanced plotting, analysis and computation capabilities for the ENDF/B-VII.0, JEFF-3.1, JENDL-3.3, ENDF/B-VI.8, ROSFOND evaluated and EXFOR experimental data libraries [1,3-7].

Sigma Web interface is an integral part of the National Nuclear Data Center (NNDC) Web Services *http://www.nndc.bnl.gov* [8] that include many complementary nuclear data resources of interest to the nuclear reactor physicists and engineers. A detailed description of Sigma interface and its selected capabilities will be presented in the following sections.

* Corresponding author

## 2. SIGMA WEB INTERFACE

Sigma's Web interface was created to extend the line of NNDC products and satisfy the data needs of the broad spectrum of users, including those who are not necessarily familiar with ENDF-6 and EXFOR formats. Its design is based on the current Web (Java, Java Server Pages, JavaScript and HTML) and relational database (MySQL and Sybase) technologies.

Sigma 1.0 was made available to the public in April 2007 and featured browsing, searching and cross sections plots for ENDF and EXFOR libraries. Version 2.0, which made its debut in April 2008, incorporated angular distribution plots and mathematical operations of cross section data. Version 3.0 was released in February 2009 and included double differential cross section plots, covariance visualization and pre-calculated thermal cross sections and resonance integrals.

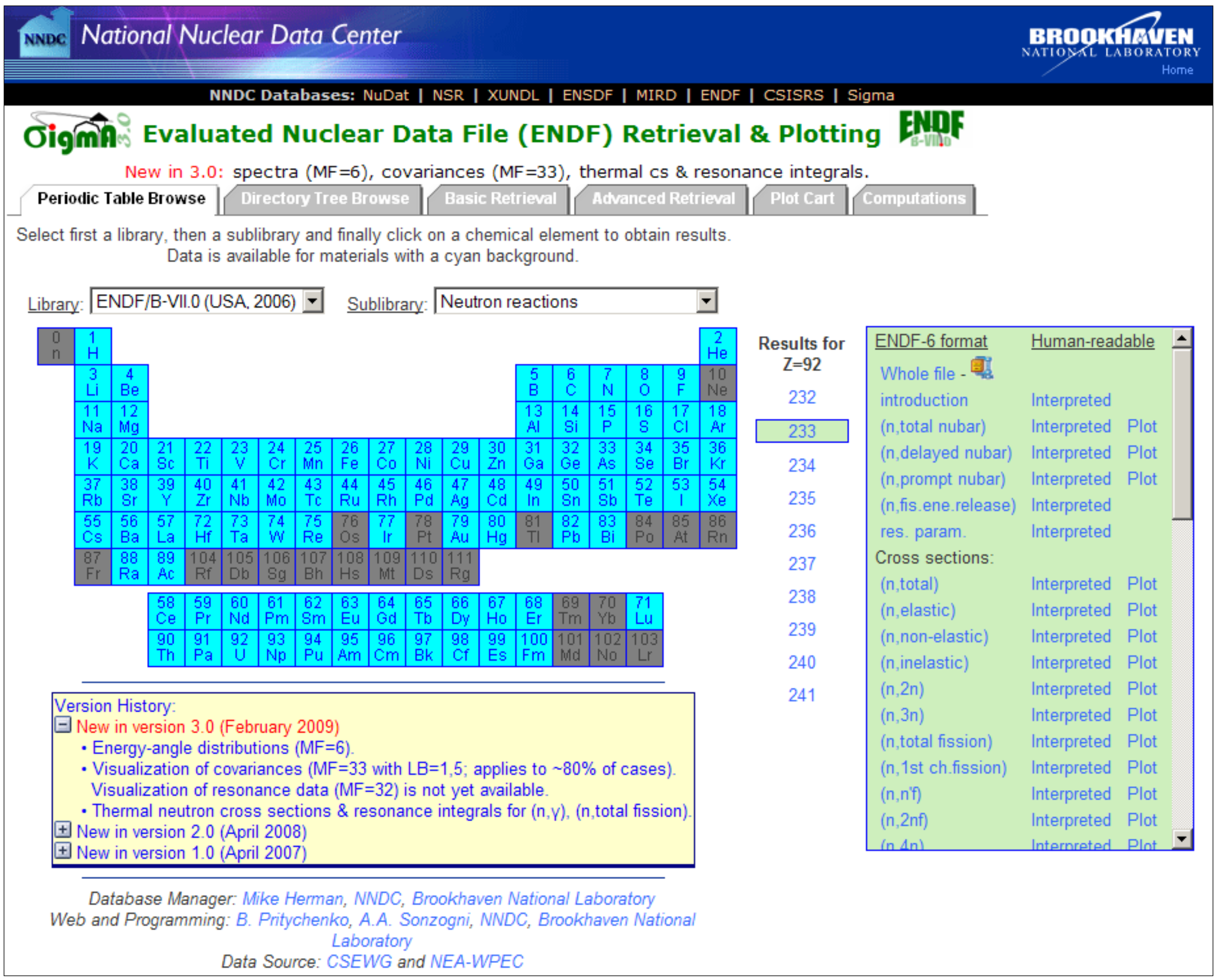

**Figure 1. Example of Sigma front page navigation capabilities (*http://www.nndc.bnl.gov/sigma*) for ENDF/B-VII.0 evaluated neutron library and $^{233}$U.**

Sigma's mission includes providing search and browsing capabilities in a transparent way for five major evaluated libraries: ENDF/B-VII.0, JEFF-3.1, JENDL-3.3, ENDF/B-VI.8 and ROSFOND [1,3-6]. The interface also offers the raw ENDF-6 data as well as processed versions and plots. Additionally, Sigma can allow the comparison between evaluated data and experimental data from the EXFOR/CSISRS database [7], and basic mathematical operations between evaluated data sets.

Interface plotting is performed by using the Java package jplots [9]. The ENDF utility codes such as PREPRO [10] are used to process the raw ENDF data and produce a point-wise version of the libraries for plotting and further computation. MySQL database server is used to store data in Sigma; a separate EXFOR database is available if necessary [11].

## 2.1. Graphical Navigation and Searching

Sigma's front page includes *Periodic Table* and *Directory Tree* options for graphical navigation and search options; on top of the *Periodic Table* tab, there are pull-down menus to select a library and sub-library. Fig.1 shows an example of what Sigma front page would look like after selecting ENDF/B-VII.0 library, neutron reactions, and clicking on Uranium (U, Z=92) and on A=233. The scrollable table on the right of the figure with a light green background shows all the data available for this material, which can be retrieved in the ENDF-6 format, and some of them processed in an interpreted form or in a plot.

Two search capabilities have been implemented, a basic one and a more advanced one. *Basic Search* shields users from having to know the intricacies of the ENDF-6 format when using Sigma. However, the *Advanced Search* feature, would require some knowledge of the format in order to make full use of it.

## 2.2. Access to Experimental Data

It is important to select high-quality ENDF evaluations before using nuclear data for reactor physics problem. Sigma Web interface provides connection to the CSISRS/EXFOR database, which contains a wealth of experimental data that has been carefully added to the database over several decades, for experimental cross sections, angular distributions and energy spectra.

The connection is made possible by the fact that both Sigma and CSISRS/EXFOR Web applications use the same *EXFOR Relational* database software [11] to facilitate the data exchange. The code X4toC4 [12] is used in Sigma to process data from CSISRS/EXFOR database for plotting. Fig. 2 shows cross sections for the $^{233}$U(n,fission) reaction from five evaluated libraries together vs. selected experimental data sets [13-15] in the $10^{-5}$ eV - 20 MeV neutron energy range.

ENDF evaluations contain angular distributions of emitted neutrons, photons and residual nuclei for a particular reaction channel. In general, many reaction channels are open and combined spectra are observed in experiments, creating a challenge when comparison between evaluation and experimental results need to be done. This problem was solved by the ENDVER code [16] that is integrated in Sigma as *Full Spectra* for $d\sigma/dE$ and $d^2\sigma/d\omega dE$ spectra.

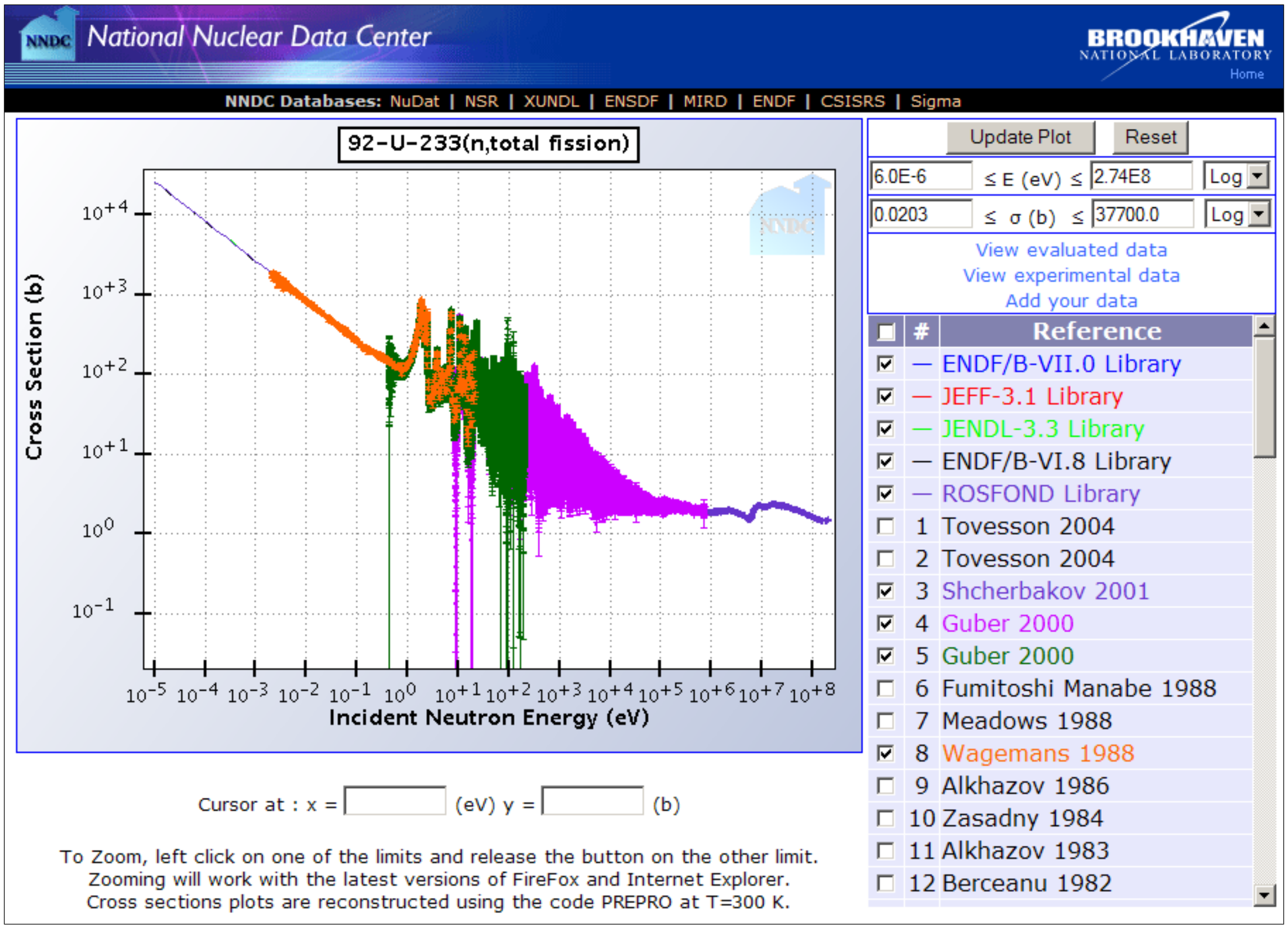

**Figure 2. Comparison between ENDF/B-VII.0, JEFF-3.1, JENDL-3.3, ENDF/B-VI.8 and ROSFOND libraries and selected experimental cross sections [13-15] for the $^{233}$U(n,fission) reaction in the $10^{-5}$ eV - 20 MeV neutron energy range.**

## 2.3. Mathematical Operations

Computations in Sigma are based on the fact that pointwise cross section values are linearized within 0.1% for a particular energy bin. This creates a possibility of calculating evaluated cross sections for any energy point within the ENDF energy range and producing a unified energy grid for neutron evaluations. The success of nuclear astrophysics calculations using ENDF libraries [17] has provided the necessary validation for this approach.

Beta version of mathematical operations in Sigma has been implemented in December of 2007. The Web application stores cross section data from the File 3 together with cross sections derived from the resonance parameters in File 2 that have been Doppler broadened to $300^0$ K and linearized using PREPRO [10]. These data can be plotted simultaneously in Sigma for a variety of different reactions (MT values) using the *Plot Cart* feature. The *Plot Cart* is made up of different sets of (x,y) data points. Almost any mathematical equation can be easily typed in using the *Computations* options and a new set of (x,y) points resulting from such a mathematical operation can be added to the *Plot Cart* and visualized.

Fig.3 shows an example of the mathematical ratio for the $^{233}$U(n,fission) cross section data from ENDF/B-VII.0 and JENDL-3.3. The disagreement between these two data sets of evaluated data in the resonance region is shown in red. Important comparisons between different data sets can be easily accomplished in Sigma.

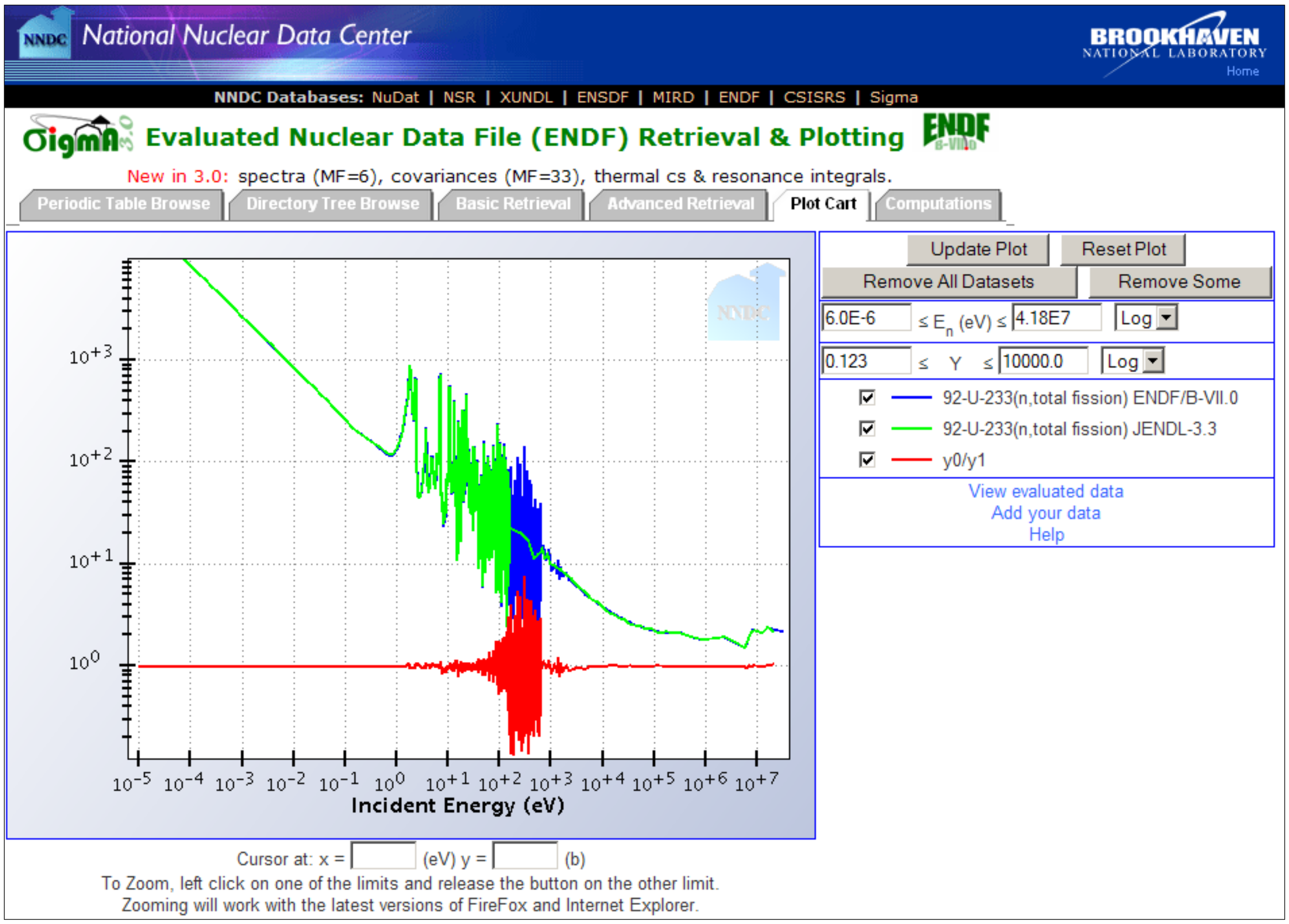

**Figure 3. The mathematical ratio (red) between ENDF/B-VII.0 (blue) and JENDL-3.3 (green) evaluated libraries for the $^{233}$U(n,fission) reaction.**

In addition to the online mathematical operations on evaluated nuclear data sets, Sigma Web interface provides pre-calculated values for thermal and 14-MeV neutron cross sections, resonance integrals and Maxwellian-averaged cross sections for 30 keV and $^{252}$Cf-fission energies that are of importance for nuclear astrophysics and reactor physics.

To address the needs of space and nuclear industries these data have been calculated for two different temperatures: 0 and 300$^{0}$ K. For quality verification current calculations have been compared with the *standards evaluation* [18] and *Atlas of Neutron Resonances* handbook [19]. ENDF/B-VII.0 calculated cross sections and resonance integrals for 0.5 eV – 20 MeV energy range are shown in Appendix A.

## 2.4. Cross Section Covariance Data

The strong interest in the production and use of covariance data is evident from the several recent sub-groups organized under the auspices of OECD-NEA as well as the current conference. In response to this demand we have developed Sigma covariance module. We have in mind a tool that would allow direct retrieval of covariance data and their visualization, which is important for both covariance producers and users. In this way we would avoid the use of complex processing codes, such as NJOY [20], where complete ENDF-6 files are needed and multigroup data must be produced before plots of uncertainties and correlations can be obtained.

In the ENDF-6 format cross section covariance matrices are stored in File 33 and the resonance parameter covariance matrices in File 32. Retrieval of File 33 is straightforward, while File 32 would need pre-processing. Direct plotting of covariance matrices shows the data exactly as given in File 33, the disadvantage being that the covariance patterns are somewhat difficult to analyze and compare.

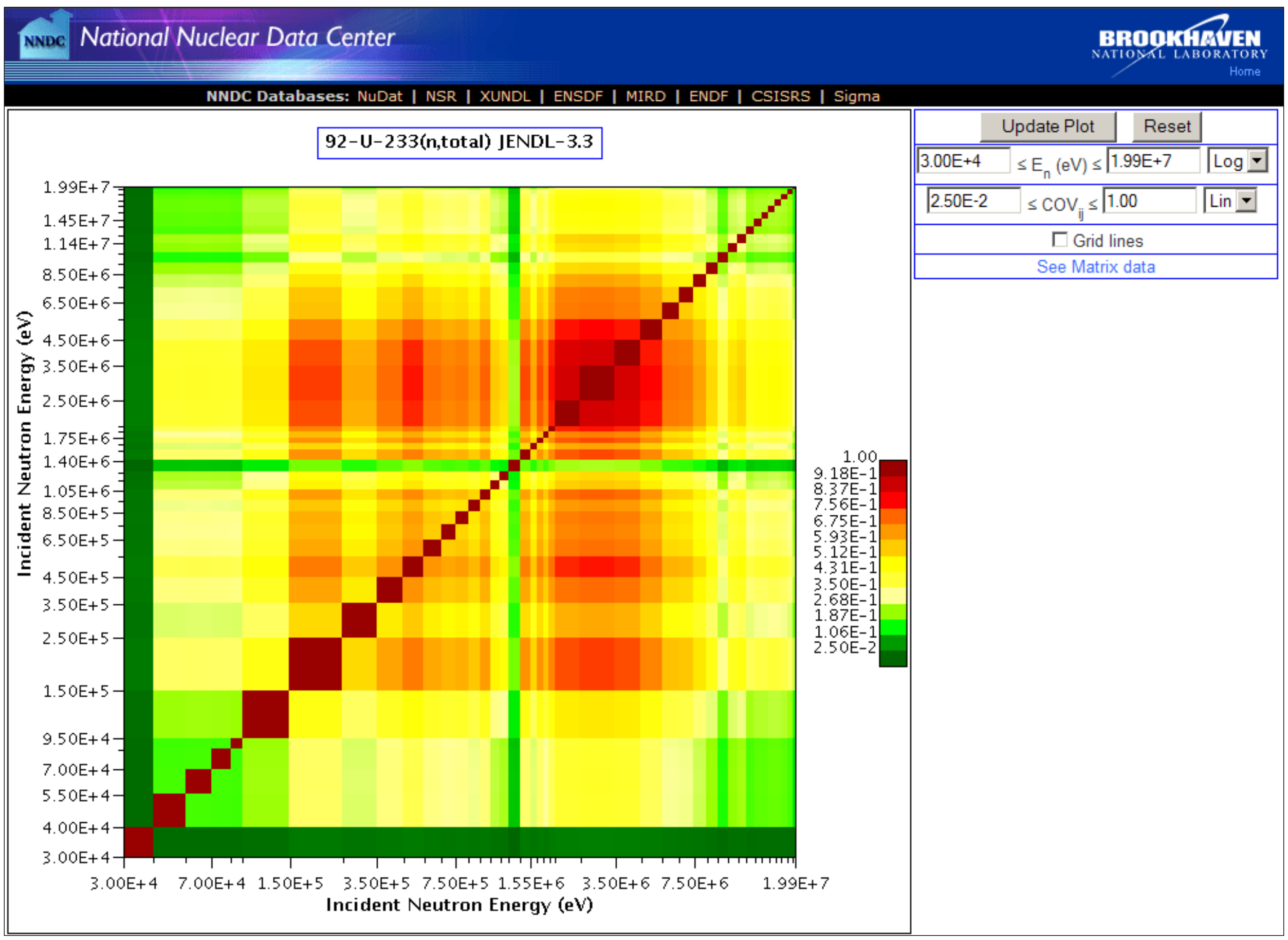

**Figure 4. JENDL-3.3 correlation matrix for the $^{233}$U(n,total) reaction.**

Often, a more suitable approach is to visualize the same data as uncertainties and correlation matrices that are obtained after simple renormalization:

$$corr(i,j) = \frac{\mathrm{cov}(i,j)}{\sqrt{\mathrm{cov}(i,i) * \mathrm{cov}(j,j)}}$$

In Fig. 4 we show an example of a correlation matrix directly retrieved from the JENDL-3.3 evaluated library for the $^{233}$U(n,total) reaction. Besides, Sigma retrievals of covariances from the preloaded libraries, such as ENDF/B-VII.0 or JENDL-3.3, Sigma Web interface technologies have been successfully extended to the *Low-fidelity Project* [21]: *http://www.nndc.bnl.gov/lowfi/.*

## 2.5. Recent Improvements

Sigma Web interface is a dynamically evolving project that constantly incorporates new features and latest computer technologies. Many changes are initiated per user requests and the interface is gaining popularity among them; it produced 144,257 and 73,908 Web retrievals in FY09 and FY08, respectively.

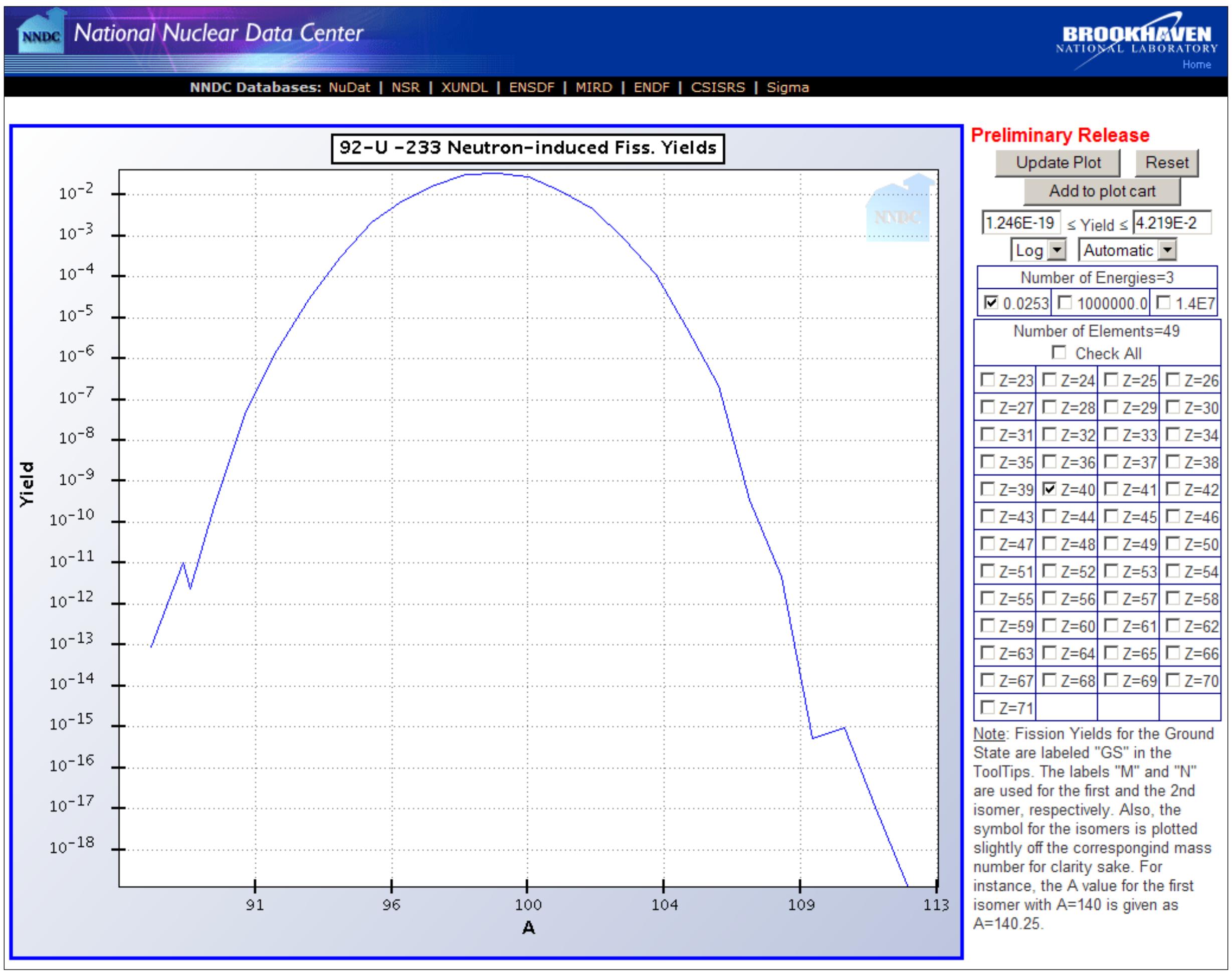

**Figure 5. JENDL-3.3 neutron-induced fission yields for $^{233}$U and Z=40.**

September 2009 developments include incorporation of 683 neutron materials from the ROSFOND library, upgrade of pre-calculated quantities, fission yields interactive plotting and upgrade of jplot zooming. Example of JENDL-3.3 neutron-induced fission yields for $^{233}$U and Z=40 is shown in Fig.5.

## 3. CONCLUSIONS

Sigma Web interface, *http://www.nndc.bnl.gov/sigma* [2], provides a user-friendly retrieval and visualization tool for the evaluated and experimental nuclear reaction data. Recent additions include energy spectra and angular distributions of emitted particles as well as mathematical operations of cross sections, fission yields and pre-calculated ENDF quantities of interest to nuclear engineers.

Sigma's covariance capabilities allow visualization of cross section covariance matrices directly retrieved from the ENDF File 33. Visualization of uncertainties and correlation matrices is straightforward and avoids the use of complex processing codes. Users will be able to retrieve and view covariance data from several preloaded libraries with a few clicks.

Five major evaluated libraries: ENDF/B-VII.0, JEFF-3.1, JENDL-3.3, ENDF/B-VI.8 and ROSFOND [1,3-6] and ENDF-A evaluations [22] are available in Sigma. In addition, users can input to Sigma their own data and view them as well. These new capabilities should not only facilitate nuclear data development, but should be appreciated also by Sigma's users. Continuing effort is needed to satisfy current needs and future demands.

## ACKNOWLEDGMENTS

We are grateful to M. Herman and A. Trkov for their help with ENDF utility codes, to P. Obložinský for useful suggestions, and to M. Blennau for a careful reading of the manuscript. This work was sponsored by the Office of Nuclear Physics, Office of Science of the U.S. Department of Energy under Contract No. DE-AC02-98CH10886 with Brookhaven Science Associates, LLC.

## APPENDIX A

Pre-calculated ENDF/B-VII.0 thermal cross sections and resonance integrals (0.5 eV – 20 MeV) values for reactor materials are shown in Tables I-II.

**Table I. ENDF/B-VII.0 pre-calculated thermal neutron cross sections and resonance integrals (0.5 eV – 20 MeV) for elastic scattering and neutron capture at T = $0^0$ K; corresponding values from the *standards evaluation* [18] and *Atlas of Neutron Resonances* [19] are shown in round and square brackets, respectively.**

| Material | $\sigma_{n,el}$ (b) | $I_{n,el}$ (b) | $\sigma_{n,\gamma}$ (b) | $I_{n,\gamma}$ (b) |
|---|---|---|---|---|
| $^{1}$H | 20.43633<br>(20.43632(*4093*))<br>[20.491(*14*)] | 263.8684 | 0.3320126<br>[0.3326(*7*)] | 0.1491554 |
| $^{7}$Li | 0.97<br>[0.97(*4*)] | 21.34454 | 0.0454<br>[0.0454(*27*)] | 0.020436<br>[0.022(*2*)] |
| $^{nat}$C | 4.73924<br>(4.73919(*948*))<br>[4.740(*5*)] | 70.53094 | 0.00336<br>[0.00350(*7*)] | 0.001688459<br>[0.00183(*5*)] |
| $^{16}$O | 3.85181<br>[3.761(*6*)] | 61.38365 | 1.9E-4<br>[0.000190(*19*)] | 8.5509E-5<br>[0.00027(*3*)] |
| $^{23}$Na | 3.319<br>[3.038(*7*)] | 131.2475 | 0.528<br>[0.517(*4*)] | 0.3164517<br>[0.311(*10*)] |
| $^{28}$Si | 1.95598<br>[1.992(*6*)] | 38.79112 | 0.1691316<br>[0.171(*3*)] | 0.08443882<br>[0.080(15)] |
| $^{40}$Ca | 3.021976<br>[2.73(*6*)] | 45.05868 | 0.4074957<br>[0.41(*2*)] | 0.2133249<br>[0.22(*2*)] |
| $^{56}$Fe | 12.05124<br>[12.62(*49*)] | 133.6497 | 2.588616<br>[2.59(*14*)] | 1.345479<br>[1.36(*15*)] |
| $^{90}$Zr | 5.481836<br>[5.3(*3*)] | 107.7711 | 0.07794076<br>[0.077(*16*)] | 0.1891033<br>[0.17(*2*)] |
| $^{113}$Cd | 26.83844 | 111.8335 | 20614.53<br>[20615(*400*)] | 391.7681<br>[390] |
| $^{197}$Au | 7.909774 | 406.941 | 98.6618 | 1570.3 |
| | [7.90(*13*)] | | (98.65931850(*13853593*))<br><br>[98.65(*9*)] | [1550(*28*)] |

**Table II. ENDF/B-VII.0 pre-calculated thermal neutron cross sections and resonance integrals (0.5 eV – 20 MeV) for elastic scattering, neutron capture and fission at T = $0^0$ K; corresponding values from the *standards evaluation* [18,1] and *Atlas of Neutron Resonances* [19] are shown in round and square brackets, respectively.**

| Material | $\sigma_{n,el}$ (b) | $I_{n,el}$ (b) | $\sigma_{n,\gamma}$ (b) | $I_{n,\gamma}$ (b) | $\sigma_{n,f}$ (b) | $I_{n,f}$ (b) |
|---|---|---|---|---|---|---|
| $^{233}$U | 12.15116<br>(12.11(*66*))<br>[12.7(*3*)] | 168.6988 | 45.23763<br>(45.56(*68*))<br>[45.5(*7*)] | 141.0432<br>[138(*6*)] | 531.2151<br>(531.22(*133*))<br>[529.1(*12*)] | 775.4899<br>[775(*17*)] |
| $^{235}$U | 15.08416<br>(14.087(*220*))<br>[14.02(*22*)] | 170.07 | 98.68643<br>(99.40(*72*))<br>[98.8(*8*)] | 140.426<br>[146(*6*)] | 585.0856<br>(584.33(*102*))<br>[582.6(*11*)] | 275.94<br>[275(*5*)] |
| $^{238}$U | 9.279782<br>[9.075(*15*)] | 346.8409 | 2.682608<br>(2.677(*12*))<br>[2.680(*19*)] | 275.5847<br>[277(*3*)] | 1.679455E-5<br>[3E-6] | 2.691588<br>[0.00163(*16*)] |
| $^{239}$Pu | 7.975233<br>(7.8(*96*))<br>[7.94(*36*)] | 178.3488 | 270.3295<br>(271.5(*214*))<br>[269.3(*29*)] | 181.3443<br>[180(*20*)] | 747.4013<br>(750(*183*))<br>[748.1(*20*)] | 302.5637<br>[303(*10*)] |
| $^{241}$Pu | 11.23797<br>(12.13(*261*))<br>[9(*1*)] | 175.1174 | 363.0489<br>(361.79(*496*))<br>[362.1(*51*)] | 179.9437<br>[162(*8*)] | 1011.852<br>(1013.96(*658*))<br>[1011.1(*62*)] | 569.6337<br>[570(*15*)] |